\documentclass[reprint,amsmath,amssymb,aps,pra]{revtex4-1}

\usepackage{graphicx}
\usepackage{dcolumn}
\usepackage{bm}

\begin{document}
\title{Comment on ``Berezinskii-Kosterlitz-Thouless  transition in two-dimensional dipolar stripes''}

\author{Fabio Cinti$^{1,2,3}$ and Massimo Boninsegni$^4$}

\affiliation{$^1$Dipartimento di Fisica e Astronomia, Universit\`a di Firenze, I-50019, Sesto Fiorentino (FI), Italy}

\affiliation{$^2$INFN, Sezione di Firenze, I-50019, Sesto Fiorentino (FI), Italy}

\affiliation{$^3$Department of Physics, University of Johannesburg, P.O. Box 524, Auckland Park 2006, South Africa}

\affiliation{$^4$ Department of Physics, University of Alberta, Edmonton, Alberta, T6G 2E1, Canada}%

\date{\today}

\begin{abstract}
In a  recent article [R. Bombin, F. Mazzanti and J. Boronat, Phys. Rev. A {\bf 100}, 063614 (2019)], it is contended that a two-dimensional  system of dipolar bosons, with dipole moments aligned at particular angles with respect to the direction perpendicular to the plane of motion, featuring a ``striped" crystalline ground state, in turn undergoes a Berezinskii-Kosterlitz-Thouless superfluid transition at low temperature, making it a two-dimensional {\em supersolid}. We show here that the results provided therein, obtained by means of Quantum Monte Carlo simulations, do {\it not} actually support such a conclusion. Rather, they are  consistent with that expounded in our work  [J. Low Temp. Phys. {\bf 196}, 413 (2019)], namely that the striped ground state is insulating (i.e., {\em non}-superfluid in the conventional sense), essentially behaving like a system of quasi-one-dimensional, parallel independent chains. We attribute the incorrectness of the conclusion reached by Bombin {\em et al.} to the very small sizes of their simulated system, which do not allow for a reliable extrapolation to the thermodynamic limit.
\end{abstract}

\maketitle
\noindent
In Ref. \onlinecite{bombin2} (henceforth referred to as BMB), the low temperature properties of a two-dimensional (2D) system of dipolar bosons, with dipole moments aligned at an angle $\alpha$  with respect to the direction perpendicular to the plane of particle motion (BMB specifically considers the case $\alpha=0.6$ rads), are studied by means of Quantum Monte Carlo (QMC) simulations at finite temperature. The main contention is that, for specific particle densities ($n$) for which the system orders in the ground state to form a ``striped'' crystal, it also undergoes a Berezinskii-Kosterlitz-Thouless (BKT) superfluid transition, making it an anisotropic 2D supersolid. 
\\ \indent 
This assertion is at variance with the outcome of a very similar study carried out by us (Ref. \onlinecite{cinti}, henceforth referred to as CB), in which we show that the striped crystal fails to develop a finite superfluid response in the transverse direction (i.e., perpendicular to that of the stripes). Specifically, at temperatures well below those at which a BKT transition would be expected, if it occurred, the value of the superfluid fraction ($\rho_S^\perp$) associated to transport in the transverse direction is {\em zero} in the limit of temperature $T\to 0$, within the statistical uncertainties of the calculation; concurrently, the one-body density matrix computed in the same direction displays a clear exponential decay. Thus, the contention was put forth that the striped crystal essentially behaves like  a collection of mostly independent quasi-one-dimensional (quasi-1D) chains, {\em not} as a genuine 2D superfluid. More generally, a full understanding of the physical behavior of the system can only be achieved by  considering {\em separately} transverse and longitudinal (i.e., along the stripes) responses.
\\ \indent 
BMB presents results for the superfluid response of the system as a function of $T$ for two different values of the density, namely $n=256$ and $n=128$, expressed in units of $a^{-2}$, $a$ being the characteristic length of the dipolar interaction (see, for instance, CB). Their estimates, obtained for different system sizes, consist of averages of the transverse and longitudinal ($\rho_S^\parallel$) superfluid fractions; they  argue that their results are indicative of a BKT transition to a superfluid phase at low $T$, for which they estimate superfluid transition temperatures. 
\\ \indent 
In actuality, though, BMB provides no evidence to the effect that the superfluid response in the transverse direction is finite, a necessary requirement if one is to establish the existence of a true 2D superfluid.
No separate estimates are given for $\rho_S^\perp$ and $\rho_S^\parallel$, only of averages of the two (ostensibly preempting the conclusion);  a vague, qualitative description is offered, suggesting that $\rho_S^\perp$ is typically small, its value being $\lesssim 0.05$ at the transition temperature, but approaches $100\%$ 
``close to the gas-stripe transition line", presumably the case for $n=128$, although this is not explicitly stated in BMB.
\hfil\break
\indent 
The system sizes on which numerical simulations were carried out in BMB are much too small, and their statistical uncertainties too large, to assess unambiguously the occurrence of a  2D BKT transition.
As we show here, the estimates of the superfluid fraction furnished in BMB, if objectively examined, are completely  consistent with the scenario proposed in CB of a non-superfluid system of independent quasi-1D chains.  The finite  superfluid response can be easily, and more plausibly attributed exclusively to the very small length of the chains.
\\ \indent 
Indeed,
the arguments furnished in BMB to rule out the possible 1D behavior of the system are spurious, as they rely on exact analytical expressions for 1D systems that only apply {\em asymptotically}, in the limit in which the system length $L \to \infty$ while the temperature $T\to 0$, with the product $LT$ held constant. They  do {\em not} yield reliable quantitative predictions for systems comprising as few as $\sim 15$ particles, which can and do display a finite, {\em large} superfluid response at low $T$, as we show below. This is, of course, a finite-size effect, but one that can be expected to contaminate significantly estimates of the overall superfluid fraction obtained on such small systems.
\\ \indent 
We begin by examining the results shown in BMB for the higher density that they consider, namely $n=256$. In this case, the systems considered by the authors comprise $N=40, 77 $ and 135 particles. We have carried out the same calculations for which results are presented in BMB, using the continuous-space Worm Algorithm, a computational methodology developed by one of us \cite{worm,worm2}, also utilized in BMB. We adopted the canonical version of it \cite{mezzacapo,mezzacapo2}, and  obtained consistent results using two separate, independent codes.\\ \indent 
Our results for the overall superfluid fraction, obtained as the average of the transverse and longitudinal parts, are in excellent quantitative agreement with those shown in Fig. 3 of BMB, within the statistical errors of both calculations; however, our physical interpretation thereof is very different. The most important observation (of course already made in CB)  is that the transverse component $\rho_S^\perp$ is {\em zero}, within the statistical uncertainties of our calculations, which we estimate to be $\lesssim 10^{-4}$. This is the case even for the smallest system size considered ($N=40$), i.e., that for which finite-size effects are most likely to yield a finite value, down to temperatures as low as one half of the lowest for which results are reported in Fig. 3 of BMB. Thus, even though there is agreement between our results for the {\em averaged} superfluid response, in our case the results unambiguously point to the superfluid signal of this system to be carried {\em entirely} along the stripes. 
\\ \indent 
In BMB, the claim is made that $\rho_S^\perp$ is finite in the $T\to 0$ limit, but the value is not provided, which makes a direct comparison of our results with theirs impossible. However, since their extrapolated ground state value of the average superfluid fraction is 0.54(5), which is  consistent  with 0.5 within statistical errors, and the computed $\rho_S^\parallel$ approaches 100\% \cite{notex}, it seems reasonable to assume that $\rho_S^\perp$ should be small -- of the order of a few percent, i.e., of the same order of magnitude of the typical statistical uncertainties quoted in BMB for $\rho_S$, at this density. Thus, we question whether the results of BMB have the precision required to resolve such a small value, and maintain that their estimates are in fact consistent with the value of zero for $\rho_S^\perp$ which we find,  supporting our physical conclusion and disproving the contention that the system is a 2D supersolid.
\\ \indent
The same remarks can be made regarding the results shown in BMB for the lower value of the density, i.e., $n=128$, for which our calculations, much like for the case mentioned above of higher density, yield a vanishing superfluid transverse component, as discussed in CB as well. Here, the situation is puzzling because, while not explicitly stated in BMB (here too, no separate estimates of $\rho_S^\perp$ and $\rho_S^\parallel$ are given,  
nor is the scaling of the superfluid fraction, displayed in Fig. 3 of BMB for $n=256$, shown in this case),
this thermodynamic point is the closest to the ``solid-gas transition line" among those investigated by the authors of BMB, and is therefore the one to which their assertion of a transverse superfluid response approaching 100\% at low $T$ should apply.\\  \indent 
However, the results for the average value of $\rho_S(T)$ featured in BMB (Fig. 4) again fail to show any statistically significant growth above 50\% at low $T$ \cite{noted}. In this case too, therefore, the results of BMB do not rule out a very small value of $\rho_S^\perp$ in the $T\to 0$ limit, nowhere near 100\%, and again, likely of the order of the statistical uncertainties quoted in BMB, i.e., again consistent with a value of zero, which is what we found in CB and in this work. 
It should be noted that a value of zero of the transverse superfluid response is entirely consistent with 
the nearly complete absence of exchanges across stripes well below the (stated) superfluid transition temperature, evident  in the top left panel of Fig. 7 of BMB, as well with 
the breaking of translational invariance in the transverse direction, which is known to be {\it incompatible} \cite{leggett} with a value of 100\% of the superfluid fraction at $T=0$. 
\\ \indent
BMB makes the claim that the results presented therein cannot be accounted for based on a picture of parallel, essentially independent quasi-1D chains (proposed in CB), because  their computed values of the superfluid fraction fail to follow the theoretically predicted 1D behavior as a function of the system size $L$ and the temperature. Specifically, they contend that their computed $\rho_S$ takes on relatively large, finite values, for system sizes and/or temperatures for which a value of zero is theoretically predicted \cite{glyde}. 
\\ \indent
As mentioned above, this argument is invalid, and can be easily refuted by simply noting that the {\em exceedingly small linear size} of the systems studied in BMB  allows for a finite superfluid response of a quasi-1D system, which can be as large as 100\% at temperatures relevant to this study. One ought not expect analytical expressions which, as explained above, are valid asymptotically, to provide reliable numerical predictions for systems comprising just a few particles. 
\\ \indent
In order to make this point more quantitatively, we  discuss results of simulations of {\em purely } 1D systems carried out in this work, aimed at modeling a single chain. We assume the same interaction among tilted dipoles as in the 2D system, and use linear densities consistent with those of BMB. 
It is stated therein that their simulations are carried out on systems enclosed in {\em rectangular} cells, and the choices of $N$ are determined by the need of simulating {\em commensurate} crystals, i.e., all $p$ stripes comprise the same number $q$ of particles. Although the actual values are not provided in BMB, the most reasonable choices for $p$ and $q$, which make the sizes of the simulation cell not too dissimilar, are 5 (6), 8 for $N=40\ (48)$, $7, 12$ for $N=84$. We assume that the larger number is always $q$, i.e., the number of particles per stripe. 
\\ \indent
A simulation of a 1D system of $N=8$ particles, of linear density $20.24$ (i.e., consistent with a 2D density equal to 256, assuming $p, q = 5, 8$) at temperature $T=128$ (the units are those adopted in BMB) yields a value of $\rho_S$  equal to 0.98(2), i.e., {\em finite, large, and in quantitative agreement} with the estimate for $\rho_S$ shown in Fig. 3 of BMB for a 2D system of $N=40$ particles at the same temperature, on assuming $\rho_S^\perp\approx 0$ and dividing by two.\\ \indent 
At lower density and temperature, the superfluid fraction of this 1D system must remain $\sim 100$\%; we have verified this to be the case for a system of linear density $14.6$ (i.e., that reported  in BMB for the case of 2D density $n=128$), and at temperature $T=72$, for which the parameter $\gamma$ used in BMB equals 2.7. For this case too, the value of the superfluid fraction is entirely  consistent with those shown in Fig. 4 of BMB for a 2D system of $N=48$ particles, assuming $p, q = 6, 8$ (and again assuming $\rho_S^\perp \approx 0$ and dividing by two). On increasing the number of particles to $N=12$ and raising the temperature to $T=60$, keeping the linear density equal to 14.6 (which means that $\gamma=3.4$), we obtain a superfluid fraction of 0.95(4), again  in agreement  with the estimates of Fig. 4 of BMB for a 2D system of $N=84$ particles (assuming $p, q = 7, 12$), within the statistical uncertainties of both calculations.
Thus, short of ``disproving'' the hypothesis of largely 1D physics, the results shown in Fig. 4 of BMB actually {\em strengthen} it. \\ \indent 
It is interesting to note that even a calculation for a system of linear density 14.6,  comprising $N=32$ particles (corresponding to  $\gamma \approx 4.5$), yields  a superfluid fraction as large as 0.74(6) at $T=30$, underscoring the importance of finite-size effects and the fallacy of applying asymptotic expressions such as that derived in Ref. \onlinecite{glyde} to systems of too small a size.
\\ \indent
The authors of BMB also show results for the {\em circularly averaged} one-body density matrix $n(r)$, whose behavior is according to them indicative of the slow power-law  decay that characterizes a BKT superfluid transition. But, as illustrated in CB, cogent insight into the physical behavior of the system is furnished {\em not} by the circularly averaged $g(r)$ but by its component in the transverse direction, which is shown to decay exponentially. On the other hand, the circularly averaged quantity for a system of size as small as even that of $N=209$ particles is strongly affected by finite-size effects coming from the longitudinal contribution.
\\ \indent
Summarizing, the study of BMB does not yield evidence of a BKT transition to a 2D supersolid phase of tilted dipolar bosons, due to the smallness of the system sizes investigated and the magnitude of their statistical uncertainties. If properly interpreted, their results are in fact consistent with the suggestion of CB, i.e., that the system displays the physical 
 behavior of an ensemble largely independent, quasi-1D chains \cite{notee}. More generally, we reiterate here our contention that no supersolid phase of dipolar bosons exists in 2D, the third dimension being required for the stabilization of such a phase \cite{cb,kora}.
 \\ \noindent
This work was supported by Natural Sciences and Engineering Research Council of Canada (NSERC). Computing support of Compute Canada is gratefully acknowledged.


\begin{thebibliography}{99}
\bibitem{bombin2}
R. Bombin, F. Mazzanti and J. Boronat, Phys. Rev. A {\bf 100}, 063614 (2019).
\bibitem{cinti}
F. Cinti and M. Boninsegni, J. Low Temp. Phys. {\bf 196}, 413 (2019).
\bibitem{worm}
M. Boninsegni, N. Prokof'ev and B. Svistunov, Phys. Rev. Lett. {\bf 96}, 070601 (2006).
\bibitem{worm2}
M. Boninsegni, N. Prokof'ev and B. Svistunov, Phys. Rev. E {\bf 74}, 036701 (2006).
\bibitem{mezzacapo}
F. Mezzacapo and M. Boninsegni, Phys. Rev. Lett. {\bf 97}, 045301 (2006).
\bibitem{mezzacapo2}
F. Mezzacapo and M. Boninsegni, Phys. Rev. A {\bf 75}, 033201 (2007).
\bibitem{notex}
This is what is invariably observed in all our calculations on  systems of this size, and is of course consistent with translational invariance in the longitudinal direction.
\bibitem{noted}
The superfluid fraction is a monotonically decreasing function of the temperature; thus, the noticeable scatter of the data shown in Fig. 4 of BMB for system size $N=48$ suggests that the statistical errors are significantly underestimated. The statistical uncertainties for greater system sizes are instead surprisingly large (approaching 20\% of the value of $\rho_S$ in some cases), making them of little practical use. 
In any case,  for all system sizes the results are consistent with an average value of the superfluid fraction saturating to 50\% in the $T\to 0$ limit (on general grounds, the superfluid fraction computed for a finite system with periodic boundary conditions can {\em only decrease} with system size).
\bibitem{leggett}
A. J. Leggett, Phys. Rev. Lett. {\bf 25}, 1543 (1970).
%
\bibitem{glyde}
L. Vranje\v{s} Marki\'c, H. Vrcan, Z. Zuhrianda, and H. R. Glyde,  Phys. Rev. B {\bf 97}, 014513 (2018).
\bibitem{notee}
One might wonder whether a regime is possible in which chains display topologically protected superfluid behavior (in the 1D sense), which could conceivably justify the denomination ``supersolid" (see, for instance, Ref. \cite{pollet}).  Our simulations do not actually support this scenario, as the value of the relevant (Luttinger) parameter is always far away from the superfluid regime, well into the quasi-crystalline one.
\bibitem{pollet}
M. Boninsegni, A. B. Kuklov, L. Pollet, N. V. Prokof'ev, B. V. Svistunov and M. Troyer, Phys. Rev. Lett. {\bf 99}, 035301 (2007). 
\bibitem{cb}
F. Cinti and M. Boninsegni, Phys. Rev. A  {\bf 96}, 013627 (2017).
\bibitem{kora}
Y. Kora and M. Boninsegni, J.  Low Temp. Phys.  {\bf 197},  337 (2019).
\end{thebibliography}
\end{document}